\documentclass[english]{report}
\usepackage[T1]{fontenc}
\usepackage[latin9]{inputenc}
\usepackage{float}
\usepackage{units}
\usepackage{amsthm}
\usepackage{amsmath}
\usepackage{amssymb}
\usepackage{graphicx}

\makeatletter

\floatstyle{ruled}
\newfloat{algorithm}{tbp}{loa}[chapter]
\providecommand{\algorithmname}{Algorithm}
\floatname{algorithm}{\protect\algorithmname}

\theoremstyle{plain}
\newtheorem{thm}{\protect\theoremname}
  \theoremstyle{definition}
  \newtheorem{defn}[thm]{\protect\definitionname}
\newenvironment{lyxcode}
{\par\begin{list}{}{
\setlength{\rightmargin}{\leftmargin}
\setlength{\listparindent}{0pt}
\raggedright
\setlength{\itemsep}{0pt}
\setlength{\parsep}{0pt}
\normalfont\ttfamily}%
 \item[]}
{\end{list}}

\@ifundefined{showcaptionsetup}{}{%
 \PassOptionsToPackage{caption=false}{subfig}}
\usepackage{subfig}
\makeatother

\usepackage{babel}
  \providecommand{\definitionname}{Definition}
\providecommand{\theoremname}{Theorem}

\begin{document}

\title{Designing Incoherent Dictionaries for Compressed Sensing: Algorithm Comparison}

\author{Eliyahu Osherovich}

\date{22/03/2009}

\maketitle

\chapter{Compressed Sensing}

\section{What is Compressed Sensing?}

Let us consider the following image which will be used extensively
throughout this introduction.

\begin{figure}
\begin{centering}
\includegraphics[width=0.4\textwidth]{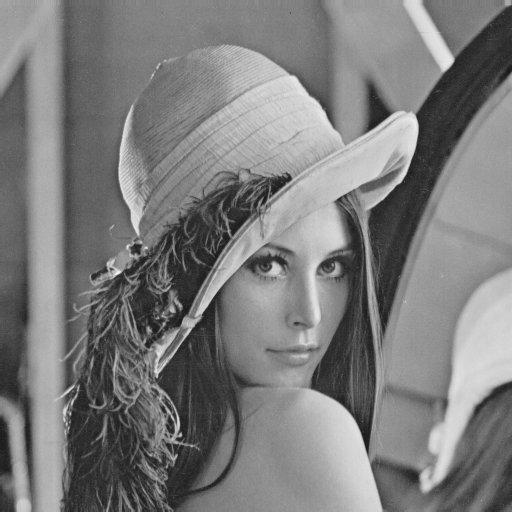}
\par\end{centering}

\caption{Lena}

\end{figure}
To insert this image into the report I pointed my editor to the file
called \texttt{Lena.jpg} which was automatically added to the text.
Let us take a closer view at the file. The size of the image is $512\times512$
pixels. Each pixel contains an integer number in the range between
0 and 255 hence, there is one byte (8 bits) of storage reserved to
keep the value of each pixel. Simple calculations show that the file
should occupy at least $512\times512\times1=262144$ bytes, however,
the size of the file is only 44791 bytes. To understand where the
difference came from let us review the way this file arrived to my
computer. First, the person portrayed in the image was photographed
by an analogous camera. Then, the film or the photograph was scanned
by a digital scanner to form a digital image that is already suitable
for use by the computer. Note that the size of this image is $512\times512$,
assuming that the original film had dimensions of $6\times6$ centimeters
we conclude that the resolution of the scanning (sampling) process
was about 85 samples per centimeter (in both directions). Sampling
rate is important if we want to reconstruct the original analog image.
Such a reconstruction is possible if we sample with the sufficiently
high rate, as defined by the Nyquist-Shannon theorem. However, we
shall return to our image which currently requires 262144 bytes of
storage. In the next step it undergoes a process called \emph{compression}.
One of the most common compression techniques for images is known
under the name JPEG compression. We describe here a very primitive
version of this compression method which, nevertheless, contains the
most important details. 

\begin{figure}
\begin{centering}
\includegraphics[width=0.4\textwidth]{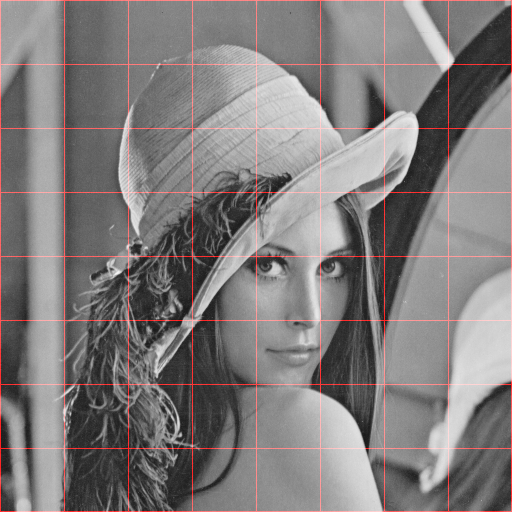}\qquad{}\includegraphics[width=0.4\textwidth]{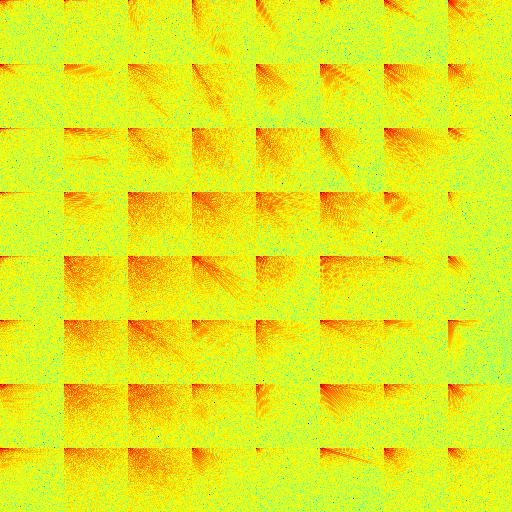}
\par\end{centering}

\caption{2D DCT on $64\times64$ patches}
\label{Flo:lena_dct}

\end{figure}

The images is splitted into square blocks and each block undergoes
the two-dimensional discrete cosine transfrom (DCT), as depicted in
Figure\ref{Flo:lena_dct}

Note that the most energy (reddish colors) is concentrated in a relatively
small fraction of the DCT coefficients, thus the coefficients that
are close to zero are discarded and we end up with a significantly
smaller number of the DCT coefficients that represent the original
image without significant visual artifacts. The original JPEG compression
does not stops here; it has a number of further steps. Those are not
important for our discussion. I would like to draw your attention
to the two important effects we achieved by discarding some of the
DCT coefficients. First, we, obviously, lost some information about
the original image, thus the compression method is \emph{lossy. }Second,
we used only a small number of the original coefficients which means
that we switched to a \emph{sparse} representation of the original
patches in the DCT base. The latter property is of paramount importance
as we shall see in the sequel. 

It turns out that sampling signals with high resolution may be wastfull
in case that the signals of interset are \emph{compressible} or, alternatively,
are known to have a \emph{sparse representation} in some basis. \emph{Compressed
Sensing} is an emerging framework dealing with efficient ways to sample
such signals.

\section{Formal definition}

\begin{figure}[H]
\begin{centering}
\subfloat[big pixel]{\begin{centering}
$\left\langle \begin{matrix}\quad\includegraphics[width=.3\textwidth]{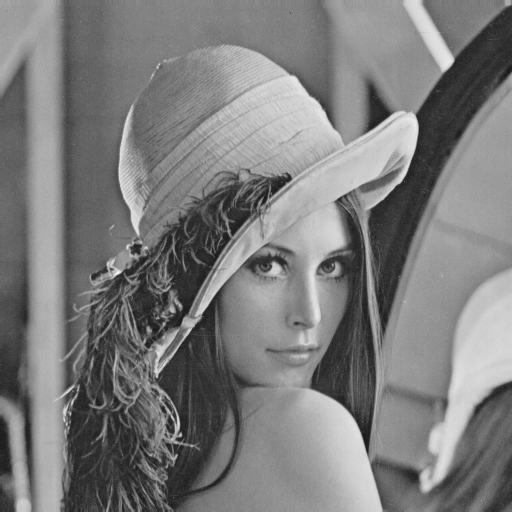} & , & \includegraphics[width=.3\textwidth]{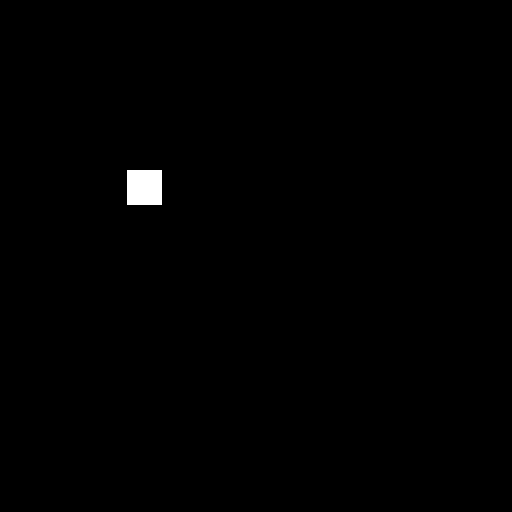}\quad\end{matrix}\right\rangle $
\par\end{centering}

}
\par\end{centering}

\begin{centering}
\subfloat[line projection (tomography)]{\begin{centering}
$\left\langle \begin{matrix}\quad\includegraphics[width=.3\textwidth]{figs/lena512} & , & \includegraphics[width=.3\textwidth]{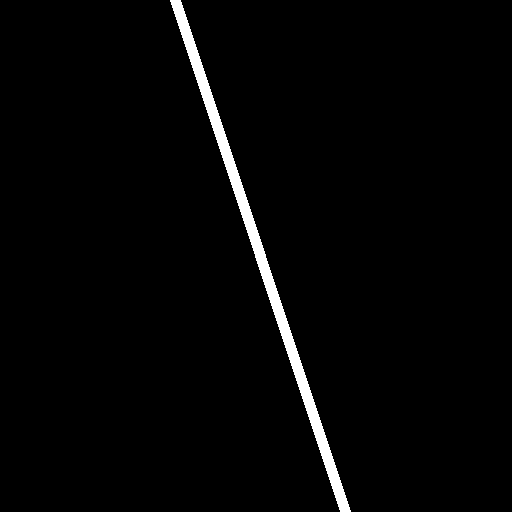}\quad\end{matrix}\right\rangle $
\par\end{centering}

}
\par\end{centering}

\begin{centering}
\subfloat[sinusoid (MRI)]{\begin{centering}
$\left\langle \begin{matrix}\quad\includegraphics[width=.3\textwidth]{figs/lena512} & , & \includegraphics[width=.3\textwidth]{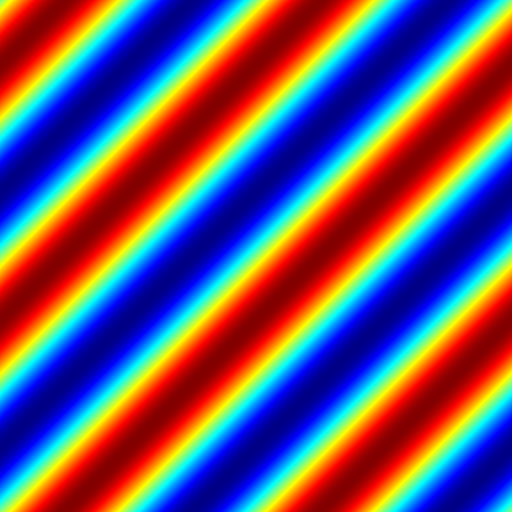}\quad\end{matrix}\right\rangle $
\par\end{centering}

}
\par\end{centering}

\begin{centering}
\subfloat[random projection]{\begin{centering}
$\left\langle \begin{matrix}\quad\includegraphics[width=.3\textwidth]{figs/lena512} & , & \includegraphics[width=.3\textwidth]{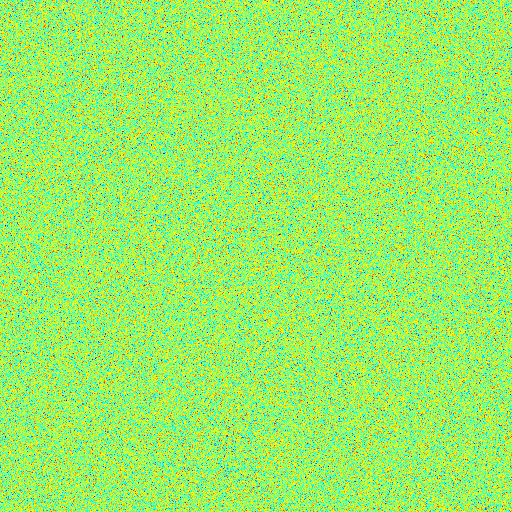}\quad\end{matrix}\right\rangle $
\par\end{centering}

}
\par\end{centering}

\centering{}\caption{Graphical representation of common projections.\label{fig:Graphical-representation-projections}}
\end{figure}

Consider a signal $x\in\mathbb{R}^{n}$ that is known to have a sparse
representation over a certain dictionary $D\in\mathbb{R}^{n\times k}$,
i.e,.
\[
x=D\theta,
\]
where $\|\theta\|_{0}\leq S\ll n$. The classical sampling methods
would require $n$ samples per signal. The Compressed Sensting, on
the other hand, suggests to replace these $n$ direct samples with
$m$ inderect ones by measuring linear projections of $x$ defined
by a projection matrix $P\in\mathbb{R}^{m\times n}$
\[
y=Px,
\]
such that $S<m\ll n$. That means that instaed of sensing $n$ elements
of the original signal we can sense it directly in compressed form,
by sampling a relatively small number of linear projections
\[
y_{i}=\left\langle p_{i},x\right\rangle .
\]
The main question with whether the original signal can be reconstructed
from this insufficient number of samples. Surprisingly, the answer
is yes. In this paper we are going to review the methods to design
efficient projections, but, meanwhile we present some of the notorious
projections used in practice in Figure~\ref{fig:Graphical-representation-projections}.

\section{Designing optimal projections}

In this section we consider the case when the dictionary $D$ is fixed
while the projection matrix $P$ can be arbitrary. Hence, we would
like to design $P$ that will suit us in the best way. Before we proceed
with the design let us define an important property that characterizes
a dictionary $D$.
\begin{defn}
For a dictionary $D$, its \emph{mutual coherence} is defined as the
largest absolute and normalized inner product between the different
columns of $D$
\[
\mu(D)=\max_{i\not=j}\frac{|d_{i}^{T}d_{j}|}{\|d_{i}\|\|d_{j}\|},
\]
where $d_{i}$ denotes the $i$-th column of $D$.

An alternative definition of the mutual coherence is via the largest
(by magnitude) off-diagonal value of the Gram matrix
\[
G=\tilde{D}^{T}\tilde{D},
\]
where $\tilde{D}$ denotes the ``normalized'' version of $D$, i.e.,
the columns of $D$ are scaled to unity (in $l_{2}$ norm). 
\end{defn}
The mutual coherence provides a quantitative measure of how far are
the columns of $D$ from being orthogonal to each other. It also has
a strong influence on the theoretical analysis of worst case scenarios,
as implied by the following theorem.
\begin{thm}
Given a signal $x=D\hat{\theta}$ such that
\[
\|\hat{\theta}\|_{0}<\frac{1}{2}\left(1+\frac{1}{\mu(D)}\right)
\]
then the following three results hold:\end{thm}
\begin{enumerate}
\item The vector $\hat{\theta}$ is necessarily the sparsest one to describe
$x$, i.e., it is the unique solution of
\[
\min_{\theta}\|\theta\|_{0}\mbox{ s.t. }x=D\theta.
\]

\item The Basis Pursuit (BP) algorithm that approximates the exact solution
$\hat{\theta}$ by solving the linear problem
\[
\min_{\theta}\|\theta\|_{1}\mbox{ s.t. }x=D\theta,
\]
is guaranteed to find the exact solution $\hat{\theta}$.
\item The Orthogonal Matching Pursuit (OMP) that is a greed approximate
algorithm which iteratively solve the least squares problem
\[
\|x-D\theta\|_{2}^{2}
\]
for only one component of $\theta$ per iteration is also guaranteed
to obtain the unique solution $\hat{\theta}$.
\end{enumerate}
With the aforementioned properties of $\mu$ there is an obvious reason
to design the projection matrix $P$ in a way that minimizes the mutual
coherence $\mu(PD)$. Note the treatment of the mutual coherence has
addressed only the worst case behavior so far. This analysis does
not provide any insight into the ``average'' behavior of the pursuit
algorithms. It is possible that relaxing the requirement on the maximal
$\mu$ value, that, on the other hand, leads to decrease in the average
$\mu$ may lead to better average performance of the pursuit algorithms.
Following this idea Elad introduced another measure which is defined
as follows.
\begin{defn}
For a dictionary $D$, its $t$\emph{-average} \emph{mutual coherence}
is defined as the average of all absolute and normalized inner products
between different columns in $D$ (denoted as $g_{ij}$) that are
above $t$. Formally it reads
\[
\mu_{t}(D)=\frac{\sum_{i\not=j}(|g_{ij}|>t)\cdot|g_{ij}|}{\sum_{i\not=j}(|g_{ij}|>t)}.
\]
For $t=0$, we obtain a simple average of the absolute entries of
$\tilde{G}$. As the value of $t$ grows, we obtain that $\mu_{t}(D)$
grows and approaches $\mu(D)$ from below. It is also evident from
the definition that $\mu_{t}(D)\geq t$. 
\end{defn}

\subsection{Elad's method}

Along with the definition given above, Elad, who followed the approach
of Dhillon \emph{et al.}, also suggested and iterative algorithm that
that tries to minimize the value $\mu_{t}(PD)$ with respect to $P$,
assuming that both the dictionary $D$ and parameter $t$ are given
and fixed the algorithm may be formalized as follows.
\begin{algorithm}[H]
\begin{lyxcode}
\textbf{Objective:}~Minimize~$\mu_{t}(PD)$~with~respect~to~$P$.

\textbf{Input:}~Use~the~following~parameters:
\begin{itemize}
\item $t$ or $t\%$ - fixed or relative threshold,
\item $D$ - the dictionary
\item $p$ - number of measurements
\item $\gamma$ - down-scaling factor
\item $Iter$ - number of iterations
\end{itemize}
\textbf{Initialization:}~set~$P_{0}\in\mathbb{R}^{m\times n}$~to~an~arbitrary~matrix~

\textbf{Loop:}~Set~$q=0$~(iteration~counter)~and~repeat~$Iter$~iterations:
\begin{enumerate}
\item \emph{Normalize: }Normalize the columns of $P_{q}D$ and obtain the
effective dictionary $\tilde{D}_{q}$.
\item \emph{Compute Gram Matrix:} $G_{q}=\tilde{D}_{q}^{T}\tilde{D}_{q}$.
\item \emph{Set Threshold:} If mode of operation is fixed, use $t$ as threshold. Otherwise,
chose $t$ such that $t\%$ of the off-diagonal elements in $G_{q}$
is above it.
\item \emph{Shrink:} Update the Gram matrix and obtain $\hat{G}_{q}$ by
\[
\hat{g}_{ij}=\begin{cases}
\gamma g_{ij} & |g_{ij}|\geq t\\
\gamma t\cdot\mathrm{sign}(g_{ij}) & t>|g_{ij}|\geq\gamma t\\
g_{ij} & \gamma t>|g_{ij}|
\end{cases}.
\]

\item \emph{Reduce Rank: }Apply SVD and force the rank of $\hat{G}_{q}$
to be equal to $m$.
\item \emph{Squared-Root: }Build the squared-root of $\hat{G}_{q}$, $S_{q}^{T}S_{q}=\hat{G}_{q}$,
where $S_{q}$ is of size $m\times k$.
\item \emph{Update $P$:} Find $P_{q+1}$ that minimizes the error $\|S_{q}-PD\|_{F}^{2}$.
\item \emph{Advance:} Set $q=q+1$.
\end{enumerate}
\end{lyxcode}
\caption{Elad's algorithm for the projection matrix $P$ optimization.}

\end{algorithm}
As we can see from the algorithm, it allows a slightly different definition
of $t$. Instead of being constant it varies from iteration to iteration
in the way that only $t\%$ of the off-diagonal entries of the Gram
matrix $G$ are altered during the current iteration.

Note that the shrinkage function used in the algorithm
\begin{equation}
\hat{g}_{ij}=\begin{cases}
\gamma g_{ij} & |g_{ij}|\geq t\\
\gamma t\cdot\mathrm{sign}(g_{ij}) & t>|g_{ij}|\geq\gamma t\\
g_{ij} & \gamma t>|g_{ij}|
\end{cases}\label{eq:elad_shrinkage}
\end{equation}
tries to preserve the ordering of the absolute entries in the Gram
matrix. Thus, the entries whose magnitude lies between $t$ and $\gamma t$
are ``shrunk'' by a smaller amount, as shown in Equation~\ref{eq:elad_shrinkage}.
This approach leads to a better distribution of the off-diagonal elements'
values, unfortunately, it also creates some large values, that were
not present in the original matrix. Large off-diagonal values ruin
completely the worst-case guarantees of the pursuit methods. The methods
presented in the next two subsections has no such undesired property.

\subsection{Duarte-Carvajalino \& Sapiro's method}

Unlike the previous method, this one is non-iterative or, more precisely,
the number of iterations is very small and constant. It was suggested
by Duarte-Carvajalino and Sapiro. Their approach is as follows. Consider
the Gram matrix of the effective dictionary $PD$:
\begin{equation}
G=D^{T}P^{T}PD,
\end{equation}
which should be as close to the identity matrix as possible, i.e.,
we would like to find such $P$ that gives approximately
\begin{equation}
D^{T}P^{T}P^{T}D\approx I.
\end{equation}
By multiplying both sides of the previous expression by $D$ on the
left and $D^{T}$ on the right we obtain
\begin{equation}
DD^{T}P^{T}P^{T}DD^{T}\approx DD^{T}.\label{eq:approx_target}
\end{equation}
Now, let us consider the singular value decomposition of $DD^{T}$
which is known, of course,
\[
DD^{T}=V\Lambda V^{T},
\]
then Equation~\ref{eq:approx_target} becomes
\begin{equation}
V\Lambda V^{T}P^{T}PV\Lambda V^{T}\approx V\Lambda V^{T},
\end{equation}
which is equivalent to 
\begin{equation}
\Lambda V^{T}P^{T}PV\Lambda\approx\Lambda.
\end{equation}
By denoting $\Gamma=PV$ we finally formulate our problem minimization
of the following functional with respect to $\Gamma$
\begin{equation}
\|\Lambda-\Lambda\Gamma^{T}\Gamma\Lambda\|_{F}^{2}.\label{eq:sapiro9}
\end{equation}
Note that if $D$ is an orthonormal basis and $m=k$, then the above
equation would have an exact solution that produces zero error, i.e.,
$\Gamma=\Lambda^{-\nicefrac{1}{2}}$. However, since the dictionary
is over-complete and $m\ll k$ we have to find an approximate solution
that minimizes the error in Equation~\ref{eq:sapiro9}. Note that
this time the error will not be equal to zero, in general. The solution
algorithm, as suggested by the authors is as follows.

Let $\lambda_{1},\lambda_{2},\ldots,\lambda_{n}$ be the singular
values of the known diagonal matrix $\Lambda$, ordered in decreasing
order, and $\Gamma=[\tau_{1}\ldots\tau_{m}]^{T}$. Then, Equation~\ref{eq:sapiro9}
becomes
\begin{equation}
\left\Vert \Lambda-\sum_{i=0}^{m}v_{i}v_{i}^{T}\right\Vert _{F}^{2},
\end{equation}
where $v_{i}=[\lambda_{i}\tau_{i,1}\ldots\lambda_{n}\tau_{i,n}]^{T}$,
or equivalently,
\begin{equation}
\left\Vert \Lambda-\sum_{i\neq j}v_{i}v_{i}^{T}-v_{j}v_{j}^{T}\right\Vert _{F}^{2}.\label{eq:sapiro10}
\end{equation}
Let us define $E=\Lambda-\sum_{i}v_{i}v_{i}^{T}$, $E_{j}=\Lambda-\sum_{i\not=j}v_{i}v_{i}^{T}$,
and let $E_{j}=U_{j}\Delta_{j}U_{j}^{T}$ be the singular value decomposition
of $E_{j}$. Then, Equation~\ref{eq:sapiro10}becomes
\[
\left\Vert E_{j}-v_{j}v_{j}^{T}\right\Vert _{F}^{2}=\left\Vert \sum_{k=1}^{n}\xi_{k,j}u_{k,j}u_{k,j}^{T}-v_{j}v_{j}^{T}\right\Vert _{F}^{2}.
\]
If we set $v_{j}=\sqrt{\xi_{1,j}}u_{1,j}$, $\xi_{1,j}$ being the
larges singular value of $E_{j}$ and $u_{1,j}$ its corresponding
eigenvector, then the largest error component in $E$ is eliminated.
Replacing $v_{j}$ back in term of $\tau_{j}$ (the rows of the matrix
we are optimizing for, $\Gamma=PV$),
\begin{equation}
[\lambda_{1}\tau_{j,1}\ldots\lambda_{n}\tau_{j,n}]^{T}=\sqrt{\xi_{1,j}}u_{1,j}.\label{eq:sapiro11}
\end{equation}
Since the matrix $\Lambda$ is in genereal not full-rank, then, for
some $r\geq0$, $\lambda_{n-r+1},\ldots,\lambda_{n}$ will be zero,
and we can only update the $\tau_{j,1},\ldots,\tau_{j,n-r}$ components
of $\tau_{j}$. This derivation forms the basis of the algorithm for
optimizing \eqref{eq:sapiro9}. The formal algorithm is given below.

\begin{algorithm}[H]
\begin{lyxcode}
\textbf{Objective:}~Minimize~$\|\Lambda-\Lambda(PV)^{T}(PV)\Lambda\|_{F}^{2}$~with~respect~to~$P$.

\textbf{Input:}~Use~the~following~parameters:
\begin{itemize}
\item $D$ - the dictionary
\end{itemize}
\textbf{Initialization:}~
\begin{enumerate}
\item find the singular value decomposition $DD^{T}=V\Lambda V^{T}$
\item set $P_{0}\in\mathbb{R}^{m\times n}$ to an arbitrary random matrix
\item set $\Gamma_{0}=P_{0}D$
\end{enumerate}
\textbf{Loop:}~Set~$q=0$~(iteration~counter)~and~repeat~$m$~iterations:
\begin{enumerate}
\item \emph{Compute $E_{q}$.}
\item Find the largest singular value and the corresponding eigenvector
of $E_{q}$, $\xi_{1,q}$and $u_{1,q}$.
\item Use \eqref{eq:sapiro11} to update the first $r$ components of $\tau_{q}$
(thereby updating $\Gamma$).
\item Compute the optimal $P=\Gamma V^{T}$.
\end{enumerate}
\textbf{EndLoop}
\end{lyxcode}
\caption{Duarte-Carvajalino \& Sapiro's algorithm for the projection matrix
$P$ optimization.}
\label{Flo:saprio_method}
\end{algorithm}

\subsection{Our method}

\begin{figure}[h]
\begin{centering}
\includegraphics{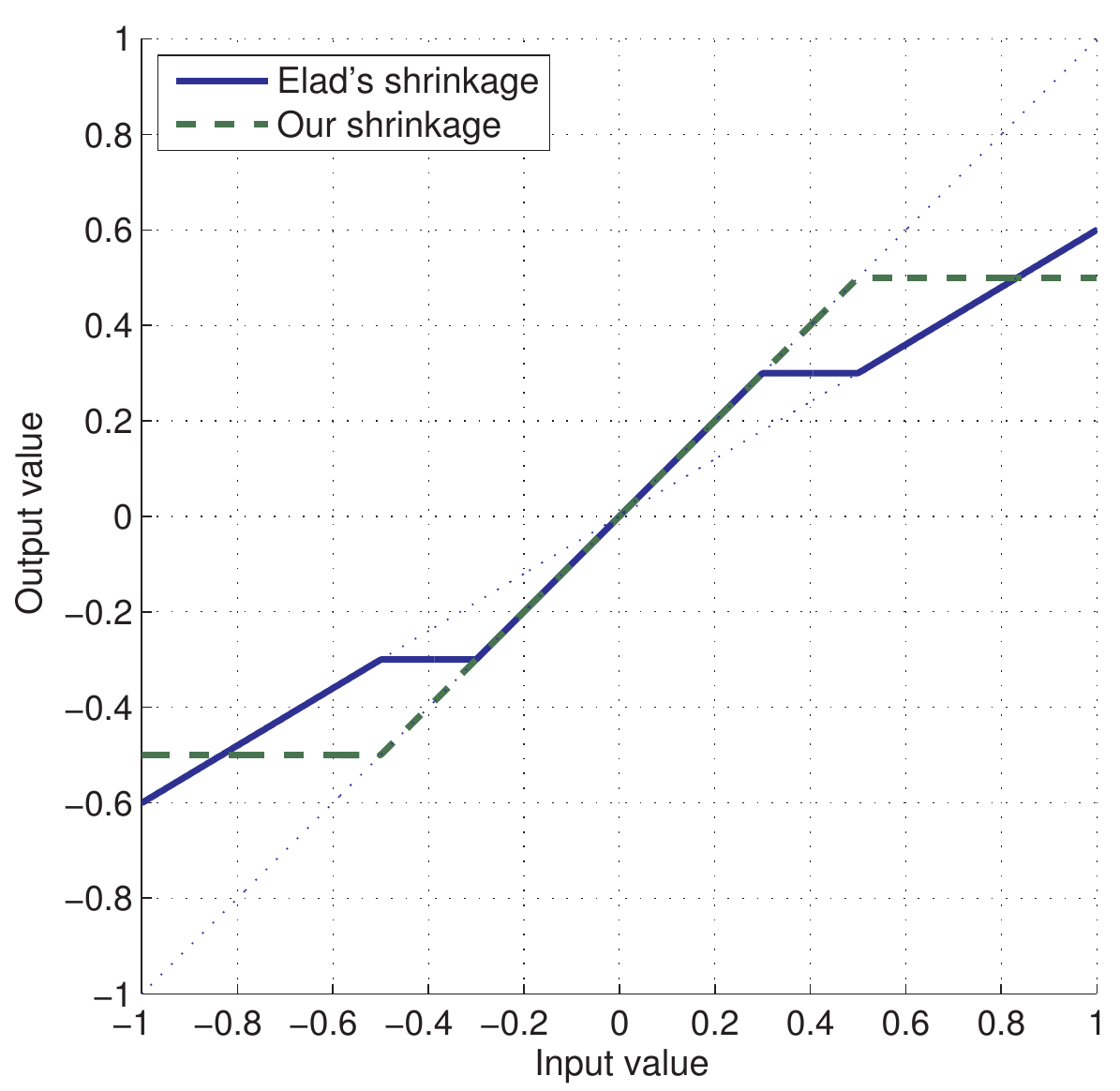}
\par\end{centering}

\caption{Shrinkage operaion in Elad's and our methods ($t=0.5$, $\gamma=0.6$)}
\label{Flo:shrinkage_visual}

\end{figure}

\begin{figure}[H]
\begin{centering}
\subfloat[Histogram of the absolute values of the off-diagonal entries of the
Gram matrix $G$ before optimization.]{\begin{centering}
\includegraphics{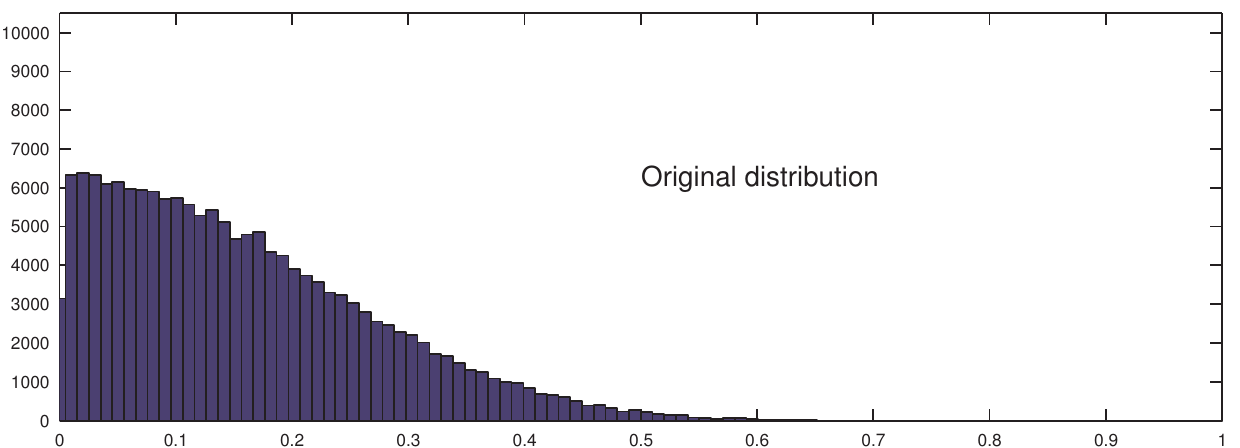}
\par\end{centering}

}
\par\end{centering}

\begin{centering}
\subfloat[Duarte-Carvajalino \& Sapiro's method]{\begin{centering}
\includegraphics{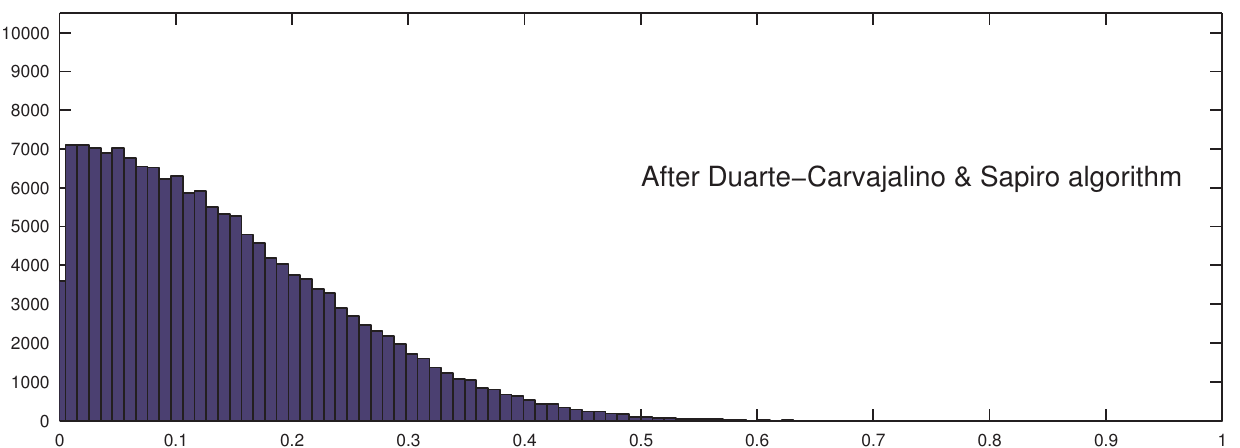}
\par\end{centering}

}
\par\end{centering}

\begin{centering}
\subfloat[Elad's method]{\begin{centering}
\includegraphics{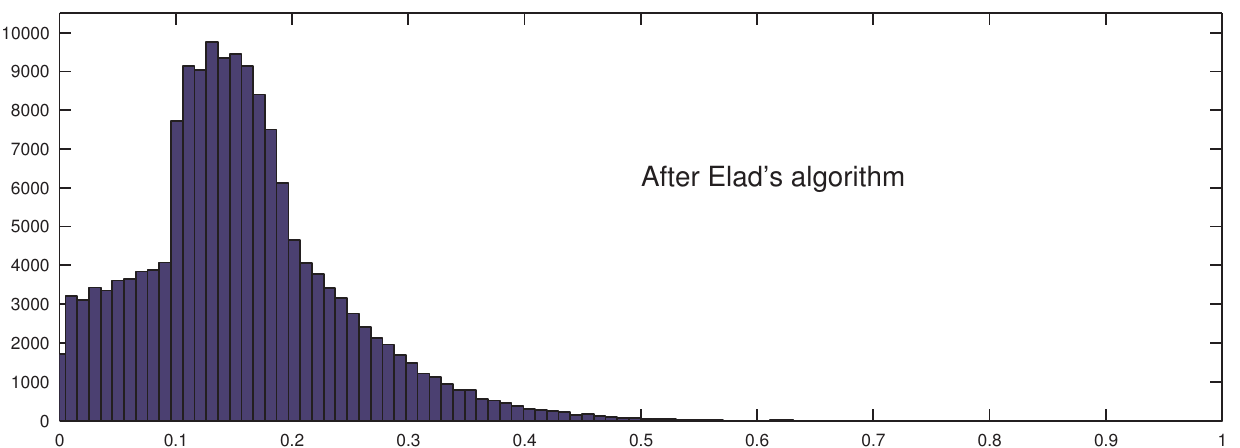}
\par\end{centering}

}
\par\end{centering}

\begin{centering}
\subfloat[Our method]{\begin{centering}
\includegraphics{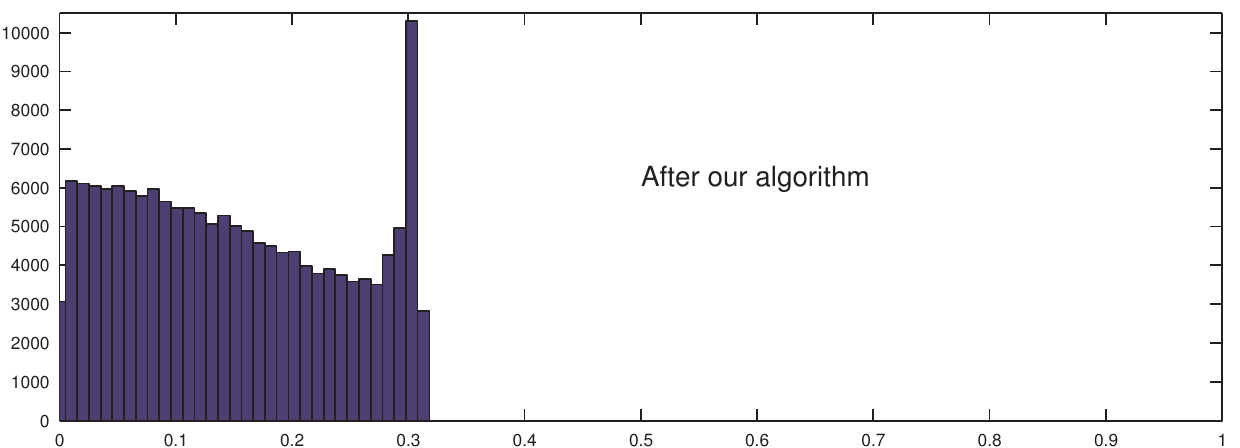}
\par\end{centering}

}
\par\end{centering}

\caption{Histogram of the absolute values of the off-diagonal entries of the
Gram matrix $G$ before the optimization and afterwards.\label{fig:histograms-after}}
\end{figure}

To overcome this drawback of the Elad's method we propose another
algorithm that models the problem in a different way. We formulate
the following feasibility problem. Find a symmetric matrix $G\in\mathbb{R}^{k\times k}$subject
to the two constraints: first, the rank of $G$ is $m<k$; second,
the off-diagonal entries must obey $|g_{ij}|\leq t$, while the diagonal
entries must be greater than or equal to unity: $|g_{ii}|\geq1$.
To solve the feasibility problem we apply the method of alternating
projections, where the current estimate $G_{q}$ is ``projected''
onto the constraint sets alternatively, as described in the algorithm
shown in Figure~\ref{Flo:our_method}.
\begin{algorithm}[H]
\begin{lyxcode}
\textbf{Objective:}~Minimize~$\mu_{t}(PD)$~with~respect~to~$P$.

\textbf{Input:}~Use~the~following~parameters:
\begin{itemize}
\item $t$ - fixed threshold,
\item $D$ - the dictionary
\item $m$ - number of measurements (rows of $P$)
\item $Iter$ - number of iterations
\end{itemize}
\textbf{Initialization:}~set~$G_{0}\in\mathbb{R}^{k\times k}$~to~an~arbitrary~random~symmetric~matrix~

\textbf{Loop:}~Set~$q=0$~(iteration~counter)~and~repeat~$Iter$~iterations:
\begin{enumerate}
\item \emph{Project onto the convex set:} Update the Gram matrix and obtain
$\hat{G}_{q}$ by

\begin{enumerate}
\item updating the off-diagonal elements $(i\not=j):$
\[
\hat{g}_{ij}=t\cdot\mathrm{sign}(g_{ij})\mbox{ if }|g_{ij}|>t
\]

\item updating the diagonal elements:
\[
\hat{g}_{ii}=1\mbox{ if }g_{ii}<1
\]

\end{enumerate}
\item \emph{Project onto the non-convex set: }force rank of $\hat{G}_{q}$
to be equal to $m$ and $\hat{G}\approx PD$.
\item \emph{Advance:} Set $G_{q+1}=\hat{G}_{q}$; $q=q+1$.
\end{enumerate}
\textbf{EndLoop}

\textbf{Return:}
\begin{enumerate}
\item \emph{Normalize} $G:$ $G=\mathrm{diag}\left(\frac{1}{\sqrt{\mathrm{diag}(G_{q})}}\right)*G_{q}*\mathrm{diag}\left(\frac{1}{\sqrt{\mathrm{diag}(G_{q})}}\right)$;
\item \emph{Find $P$:} solve $D^{T}P^{T}PD=G$ for $P$.
\end{enumerate}
\end{lyxcode}
\caption{Our algorithm for the projection matrix $P$ optimization.}
\label{Flo:our_method}
\end{algorithm}
 The main difference between the two algorithm is the ``shrinkage''
methods, which are shown graphically in Figure\ref{Flo:shrinkage_visual},
and absence of additional parameter $\gamma$ in our method. 
 Our method does not produce large off-diagonal elements, in fact,
it succeeds to remove all elements that are larger than the given
threshold $t$, provided that the target was not too aggressive. Effect
of applying both algorithm to the distribution of the magnitude of
the off-diagonal entries is demonstrated in Figure~\ref{fig:histograms-after}.
In this experiment we used a random dictionary $D$ of size $200\times400$.
Its entries we randomly drawn from a normal distribution with zero
mean and unity variance. Number of the projections were set to 30,
i.e., the size of the projection matrix was $30\times200$. Effective
threshold we used was $26\%$ for the Elad's algorithm and $0.26$
for our method. The parameter $\gamma$ was set to 0.6 in the Elad's
algorithm.\pagebreak{}\thispagestyle{empty}
\pagebreak{}

\begin{figure}[H]
\begin{centering}
\subfloat[Original]{\begin{centering}
\includegraphics[width=0.9\textwidth]{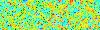}
\par\end{centering}

}
\par\end{centering}

\begin{centering}
\subfloat[After Duarte-Carvajalino \& Sapiro's method]{\begin{centering}
\includegraphics[width=0.9\textwidth]{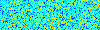}
\par\end{centering}

}
\par\end{centering}

\begin{centering}
\subfloat[After Elad's method]{\begin{centering}
\includegraphics[width=0.9\textwidth]{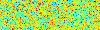}
\par\end{centering}

}
\par\end{centering}

\begin{centering}
\subfloat[After our method]{\begin{centering}
\includegraphics[width=0.9\textwidth]{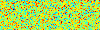}
\par\end{centering}

}
\par\end{centering}

\caption{Projection matrix after optimization (a subset of 100 columns).\label{fig:Projection-matrix-after-structure}}
\end{figure}

After the optimization, the projection matrix $P$ is both cases does
not reveal any specific structure, as we can see in Figure~\eqref{fig:Projection-matrix-after-structure}
\pagebreak{}\thispagestyle{empty}
\begin{figure}[H]
\begin{centering}
\subfloat[Initial random projection matrix $P$]{\begin{centering}
\includegraphics{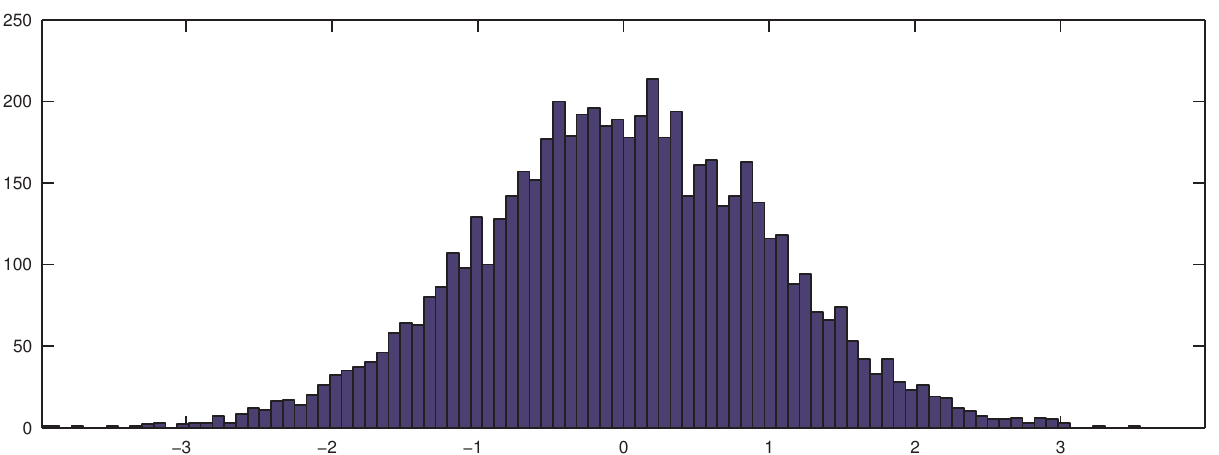}
\par\end{centering}

}
\par\end{centering}

\begin{centering}
\subfloat[After Duarte-Carvajalino \& Sapiro's method]{\begin{centering}
\includegraphics{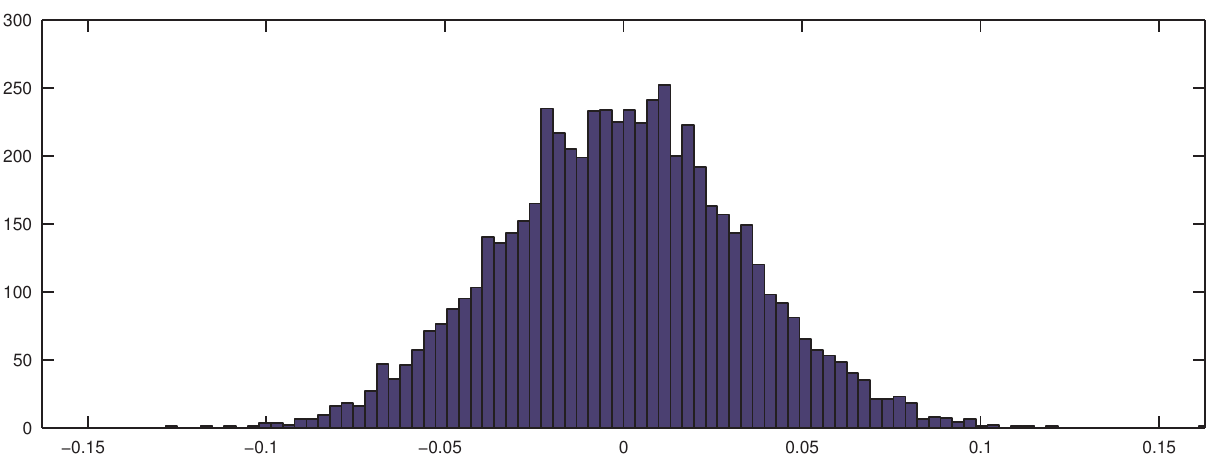}
\par\end{centering}

}
\par\end{centering}

\begin{centering}
\subfloat[After Elad's method]{\begin{centering}
\includegraphics{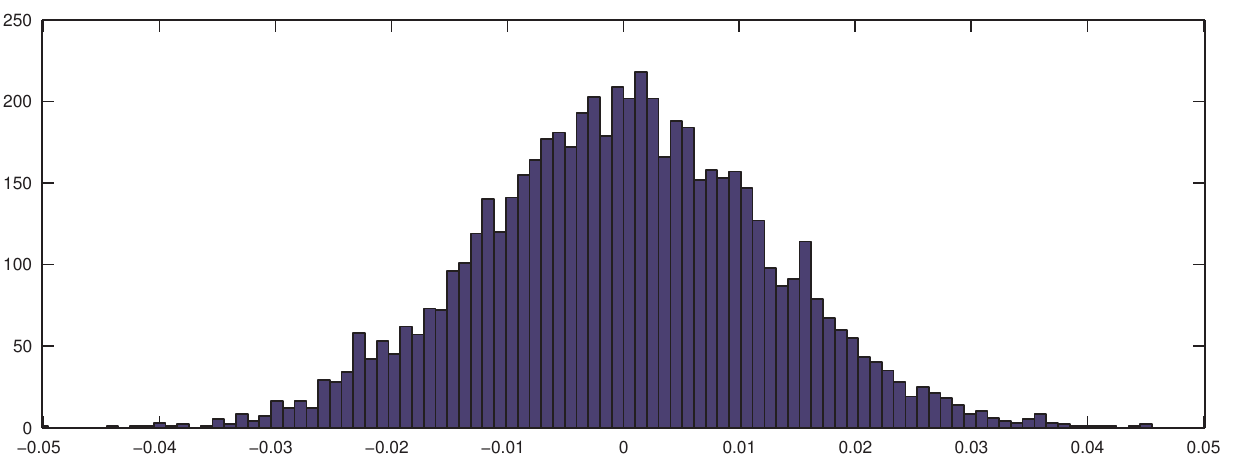}
\par\end{centering}

}
\par\end{centering}

\begin{centering}
\subfloat[After our method]{\begin{centering}
\includegraphics{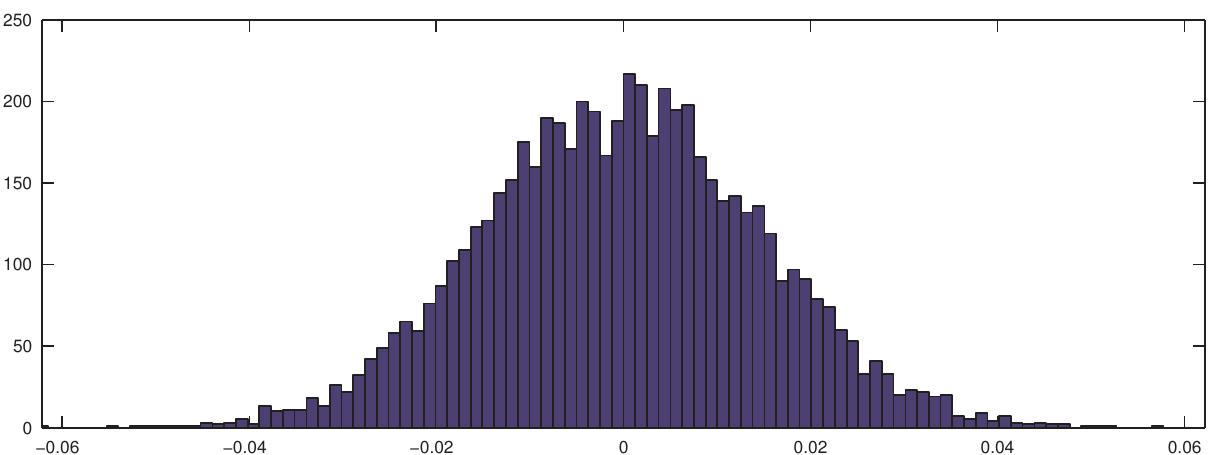}
\par\end{centering}

}
\par\end{centering}

\caption{Histogram of the values of the projection matrix $P$ before the optimization
and afterwards.\label{fig:Projection-matrix-after-values}}
\end{figure}
\pagebreak{}

It is also interesting to look at the distribution of the values in
the projection matrix before the optimization and afterwards. It seems
that the values are still drawn form the Gaussian distribution with
zero mean, but the variance is much lower. Our experiments indicate
that this is a typical result, that does not depend on the initial
guess. 

\begin{figure}[H]
\begin{centering}
\includegraphics{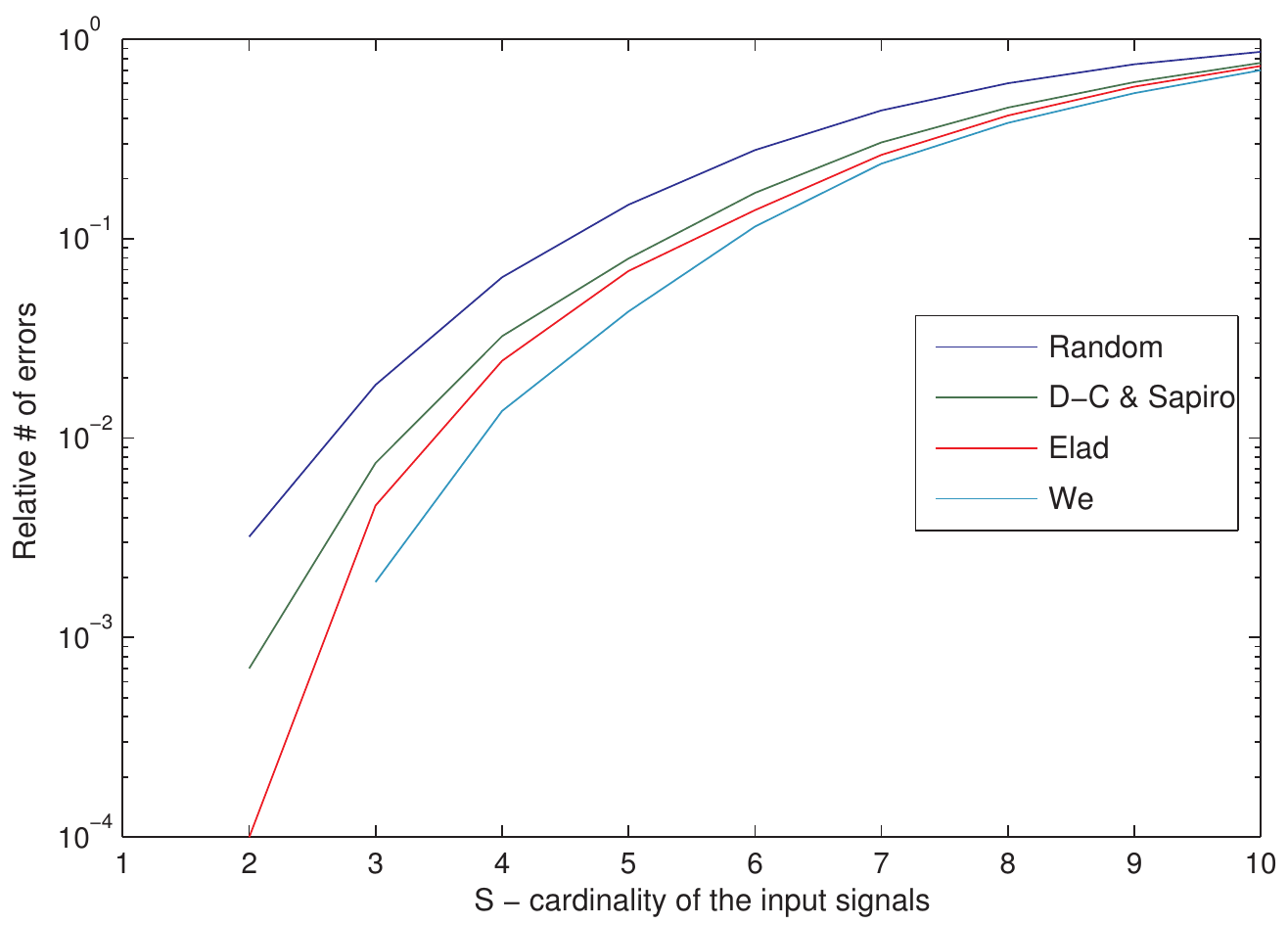}
\par\end{centering}

\caption{CS reconstruction relative errors as a function of the signals' cardinality
$S$. Using OMP method.\label{fig:omp-performance}}

\end{figure}

\begin{figure}[H]
\begin{centering}
\includegraphics{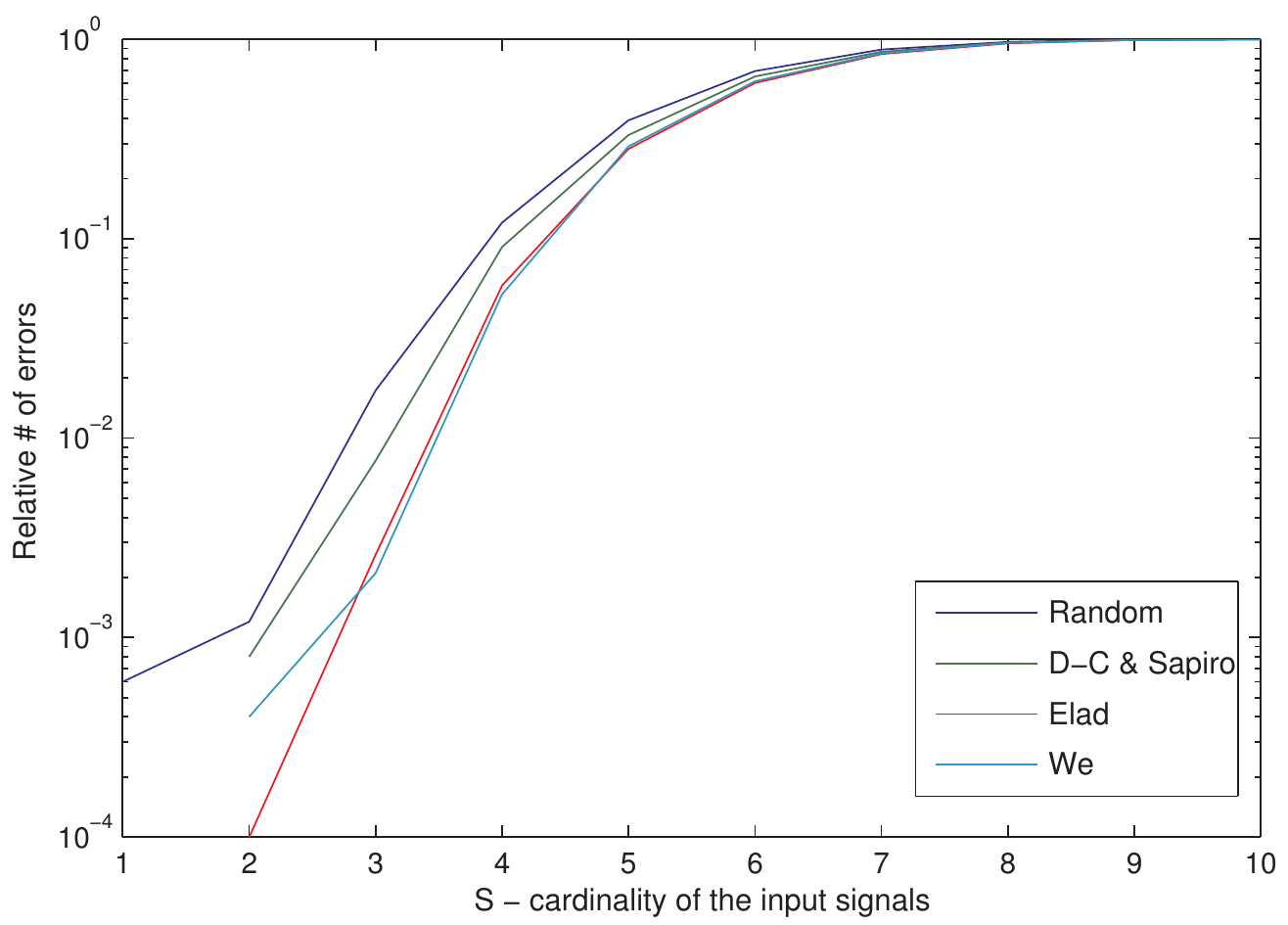}
\par\end{centering}

\caption{CS reconstruction relative errors as a function of the signals' cardinality
$S$. Using BP method.\label{fig:BP-perfomance}}
\end{figure}

If we consider the final distribution of the off-diagonal entries
of the Gram matrix we will find out that it is quite different, as
is evident from Figure~\ref{fig:histograms-after}. The correlation
between the optimized projections matrix $P$ is yet to be theoretically
analyzed, therefore, to evaluate the performance of the compressed
sensing with and without the optimized projections we performed the
following tests:
\begin{description}
\item [{Stage~1:}] Generate data: choose a dictionary $D\in\mathbb{R}^{n\times k}$
and synthesize $N$ test signals $\{x_{i}\}_{i=1}^{N}$ by generating
$N$ sparse vectors $\{\theta_{i}\}_{i=1}^{N}$ of length $k$ each,
and computing $x_{i}=D\theta_{i}$ for all $i$. All representations
$\{\theta_{i}\}$ are built with the same low cardinality $\|c_{i}\|=S$.
\item [{Stage~2:}] Random projections: for a chosen number of projections
$m$ generate a random projection matrix $P\in\mathbb{R}^{m\times n}$
and apply it to the signals to obtain $y_{i}=Px_{i}$. Compute the
effective dictionary $\hat{D}=PD$.
\item [{Stage~3:}] Performance tests: apply the BP and OMP to reconstruct
the signals by approximate the solution of
\[
\hat{\theta}_{i}=\arg\min_{\theta}\|\theta\|_{0}\mbox{ s.t. }y_{i}=\hat{D}\theta_{i}.
\]
The result obtained by the pursuit algorithms is compared against
the true solution by evaluating the error $\|\hat{\theta}_{i}-\theta_{i}\|$.
Errors above some threshold are considered as a reconstruction failure.
\item [{Stage~4:}] Optimized projections: the evaluations above are repeated
for the projection matrix as returned by the optimization algorithms.
\end{description}
We have followed the above stages to evaluate how the influence of
the projection matrix optimization varies along with the cardinality
of the true solution. In our experiment we used a random dictionary
of size $200\times400$, i.e., $n=200$, $k=400$. For different of
$S$ ($S=1,2,\ldots,10$) we generated $N=10000$ sparse vectors of
length $k=400$ with $S$ non-zero values in each. The locations of
the non-zero components were chosen at random and populated with i.i.d.
zero-mean and unit variance Gaussian values. These sparse vectors
were used to create the example signals that were used in the evaluation
of the CS performance. The results for the OMP and BP are depicted
in Figures~\ref{fig:omp-performance} and~\ref{fig:BP-perfomance}
respectively. 

\section{Simultaneous optimization of the projection matrix and the dictionary}

The idea of designing an optimal projection matrix can be taken further
if we allow the dictionary $D$ to be changed as well. Let us recall
the K-SVD algorithm that is is used for dictionary training. Given
an $n\times p$ matrix $X=[x_{1},x_{2},\ldots,x_{p}]$ of $p$ training
images of length $n$ pixels each, used to train on overcomplete dictionary
$D$ of size $n\times k$, with $p\gg k$ and $k>n$. The objective
of the K-SVD algorithm is to solve, for a given sparsity level,
\begin{equation}
\min_{D,\Theta}\|X-D\Theta\|_{F}^{2}\mbox{ s.t. }\|\theta_{i}\|_{0}\leq S,\label{eq:sapiro_13}
\end{equation}
where $\Theta=[\theta_{1},\theta_{2},\ldots,\theta_{p}]$, and $\theta_{i}$
is the sparse vector of coefficients representing the $i$-th image
in terms of columns of the dictionary $D=[d_{1},d_{2},\ldots,d_{k}]$.
K-SVD is an iterative algorithm that progressively improves the functional
in Equation~\eqref{eq:sapiro_13} as described next. First, the algorithm
freezes the current dictionary $D$ and solves for the coefficients'
matrix $\Theta$ using one of the pursuit algorithms. Next, it assumes
that the matrises $\Theta$ and $D$ are fixed except for one column
of $D$: $d_{j}$. Finding the best $d_{j}$ is done by re-writing
the functional in Equation~\eqref{eq:sapiro_13} as follows
\begin{equation}
\left\Vert X-D\Theta\right\Vert _{F}^{2}=\left\Vert \left(X-\sum_{i\not=j}d_{i}\theta_{T}^{j}\right)-d_{j}\theta_{T}^{j}\right\Vert _{F}^{2}=\left\Vert E_{j}-d_{j}\theta_{T}^{j}\right\Vert _{F}^{2},\label{eq:ksvd}
\end{equation}
where $\theta_{T}^{j}$ denotes the $j$-th row of $\Theta$. It is
tempting to use the SVD to solve for new $d_{j}$ and $\theta_{T}^{j}$,
however, that will lead to a dense $\theta_{T}^{j}$ which is absolutely
undesirable. The K-SVD algorithm overcomes this difficulty by using
a small subset of columns of $X$ that use the atom $d_{j}$ in their
representation. Consequently, the only updated entries of $\theta_{T}^{j}$
are those that are non-zero at the moment. Hence, the algorithm does
not increase the density of the coefficients matrix $\Theta.$ After
running over all columns of $D$ the algorithm returns to the first
step and cycles until convergence, or maximum number of iterations.
There exists experimental evidence that the K-SVD algorithm is sensitive
to the initial guess, however, this question is not addressed here. 

The Coupled-KSVD algorithm suggested by Duarte-Carvajalino and Sapiro
extends the KSVD for simultaneous training of the projection matrix
$P$ and dictionary $D$ with images available from a dataset. They
define the following objective function
\begin{equation}
\min_{P,D,\Theta}\lambda\left\Vert X-D\Theta\right\Vert _{F}^{2}+\left\Vert Y-PD\Theta\right\Vert _{F}^{2}\mbox{ s.t. }\|\theta_{i}\|_{0}\leq S,\label{eq:sapiro_19}
\end{equation}
where the scalar $\lambda\in[0,1]$ controls the relative weight of
the two terms and $Y$ are linear samples given by
\begin{equation}
Y=PX+N,
\end{equation}
where $N$ represents an additive noise introduced by the sensing
system. Let us denote
\begin{equation}
Z=\left(\begin{array}{c}
\begin{matrix}\lambda X\\
Y
\end{matrix}\end{array}\right),\, W=\left(\begin{matrix}\lambda I\\
P
\end{matrix}\right).
\end{equation}
Then, Equation~\eqref{eq:sapiro_19} can be re-written as,
\[
\min_{P,D,\Theta}\left\Vert Z-\tilde{D}\Theta\right\Vert \mbox{ s.t. }\left\Vert \theta_{i}\right\Vert _{0}\leq S,
\]
where $\tilde{D}=WD$. Now, in exactly the same manner as we did in
the K-SVD we run over all columns of $\tilde{D}$ one by one and find
simultaneous solution for $\tilde{d}_{j}$ and $\theta_{T}^{j}$.
Then, we substitute back
\begin{equation}
\tilde{d}_{j}=\left(\begin{matrix}\lambda I\\
P
\end{matrix}\right)d_{j}.\label{eq:sapiro_28}
\end{equation}
At this point, we have $m+n$ equations and $n$ unknowns. The solution
is obtained by solving the overdetermined system in the sense of the
least squares, thus the unknown column $d_{j}$ can be found by
\begin{equation}
d_{j}=\left(\lambda^{2}I+P^{T}P\right)^{-1}\left(\begin{matrix}\lambda I & P^{T}\end{matrix}\right)\tilde{d}_{j}.\label{eq:sapiro_29}
\end{equation}
Finally, the norm of $d_{j}$ is adjusted to unity (of course, this
step must be accompanied by adjusting the row $\theta_{T}^{j}$) to
keep the product $\tilde{d}_{j}\theta_{T}^{j}$ without any change),
i.e.,
\begin{equation}
\theta_{T}^{j}=\theta_{T}^{j}\|d_{j}\|_{2},\quad d_{j}=d_{j}/\|d_{j}\|_{2}\label{eq:sapiro_30}
\end{equation}
After we finished the update of the dictionary $D$ and the coefficient
matrix $\Theta$, assuming fixed $P$, we can update the projection
matrix, with the methods described in the previous section, i.e.,
learning the projection matrix from the just updated dictionary $D$.
Then, we repeat the algorithm until convergence. Hence, the whole
algorithm is formalized in Algorithm~\ref{alg:Coupled-KSVD}.
\begin{algorithm}
\begin{lyxcode}
\textbf{Objective:}~Minimize~$\lambda\left\Vert X-D\Theta\right\Vert _{F}^{2}+\left\Vert Y-PD\Theta\right\Vert _{F}^{2}\mbox{ s.t. }\|\theta_{i}\|_{0}\leq S$~with~respect~to~$P$,$D$,~and~$\Theta$.

\textbf{Input:}~Use~the~following~parameters:
\begin{itemize}
\item $X$,$Y$ - the training data and its projections
\item $\lambda$ - weight of the first term ($0\leq\lambda\leq1$)
\end{itemize}
\textbf{Initialization:}~
\begin{enumerate}
\item set initial dictionary $D_{0}\in\mathbb{R}^{p\times n}$ to an arbitrary
random matrix
\end{enumerate}
\textbf{Loop:}~Set~$q=0$~(iteration~counter)~and~repeat~until~convergence:
\begin{enumerate}
\item For fixed $D_{q}$ compute $P_{q}$ using an algorithm from the previous
section.
\item For fixed $D_{q}$ and $P_{q}$ compute $\Theta_{q}$ using a pursuit
algorithm, e.g., OMP.
\item Using update $P_{q}$, $D_{q}$, and $\Theta_{q}$ by the K-SVD, as
described in Equations~\ref{eq:sapiro_28},\ref{eq:sapiro_29}, and
\ref{eq:sapiro_30}. 
\end{enumerate}
\textbf{EndLoop}
\end{lyxcode}
\caption{Coupled KSVD.\label{alg:Coupled-KSVD}}

\end{algorithm}
Experimental results of this approach are available in the original
paper.

\section{Conclusions}

We presented several algorithms for optimization of the projection
matrix used in the Compressed Sensing. Experimental results indicate
that our method performs better than the the other two. Simultaneous
method of optimization of the projection matrix and the sparsifying
dictionary as suggested by Duarte-Carvajalino and Sapiro was presented
in a descriptive manner without experimental results.

\bibliographystyle{plain}
\nocite{*}
\bibliography{/home/eli/technion/236601-Miki/project/biblio/Bibliography}

\begin{thebibliography}{1}

\bibitem{aharon_k-svd:algorithm_2006}
M.~Aharon, M.~Elad, and A.~Bruckstein.
\newblock {K-SVD:} an algorithm for designing overcomplete dictionaries for
  sparse representation.
\newblock {\em {IEEE} Transactions on signal processing}, 54(11):4311, 2006.

\bibitem{duarte-carvajalino_ima_????}
J.~M. {Duarte-Carvajalino} and G.~Sapiro.
\newblock {IMA} preprint series\# 2211.

\bibitem{elad_optimized_2007}
M.~Elad.
\newblock Optimized projections for compressed sensing.
\newblock {\em Signal Processing, {IEEE} Transactions on}, 55(12):5695--5702,
  2007.

\bibitem{tropp_designing_2005}
J.~A. Tropp, I.~S. Dhillon, R.~W. Heath, and T.~Strohmer.
\newblock Designing structured tight frames via an alternating projection
  method.
\newblock {\em {IEEE} Transactions on information theory}, 51(1):188--209,
  2005.

\end{thebibliography}

\end{document}